\documentclass[final,1p,times]{elsarticle} 
\usepackage{graphicx} 
\usepackage{amssymb} 
\usepackage{amsthm} 

\journal{Nuclear Physics A} 
\begin{document} 

\begin{frontmatter} 

\title{Highlights from PHENIX II - Exploring the QCD medium}

\author{Carla M. Vale$^{a}$ for the PHENIX collaboration}

\address[a]{Physics Department,  
Brookhaven National Laboratory,
Upton, NY, 11973, USA}

\begin{abstract} 
Much of the present experimental effort at RHIC is now directed
towards understanding the properties of the hot and dense colored
medium created in A+A collisions. Recent results from PHENIX on the
dynamical evolution of the medium and its response to high momentum
probes are presented, and their impact on our overall
understanding of heavy-ion collisions is discussed.
\end{abstract} 

\end{frontmatter} 

\section{Introduction}

Collisions of heavy nuclei at relativistic energies, such as the ones at RHIC, provide a unique opportunity for the study of nuclear matter at very high temperatures and energy densities, similar to those that existed in the earliest moments of the universe. Under these conditions matter is expected to enter a state where quarks and gluons are no longer confined into hadrons. The motivation for the experimental program at RHIC is to explore this transition and study the QCD medium that evolves from it. 

The first few years of the RHIC experimental program provided significant insight into the general properties of the medium produced in RHIC collisions, and established that it is hot, dense and strongly coupled, behaving like a near-perfect fluid (for a review, see \cite{bib12}, also previous volumes in the series of Quark Matter proceedings).  More recently, with the advent of larger data sets and collision energy scans, it is becoming possible to analyze the data in a more quantitative fashion. With the help of theoretical models, we aim to extract physical quantities from the data, in order to characterize the matter produced at RHIC and its response to the probes used in testing it. 

\section{Azimuthal Anisotropy and Scaling Properties}

\begin{figure}[t]
\centering
\includegraphics[width=0.85\textwidth]{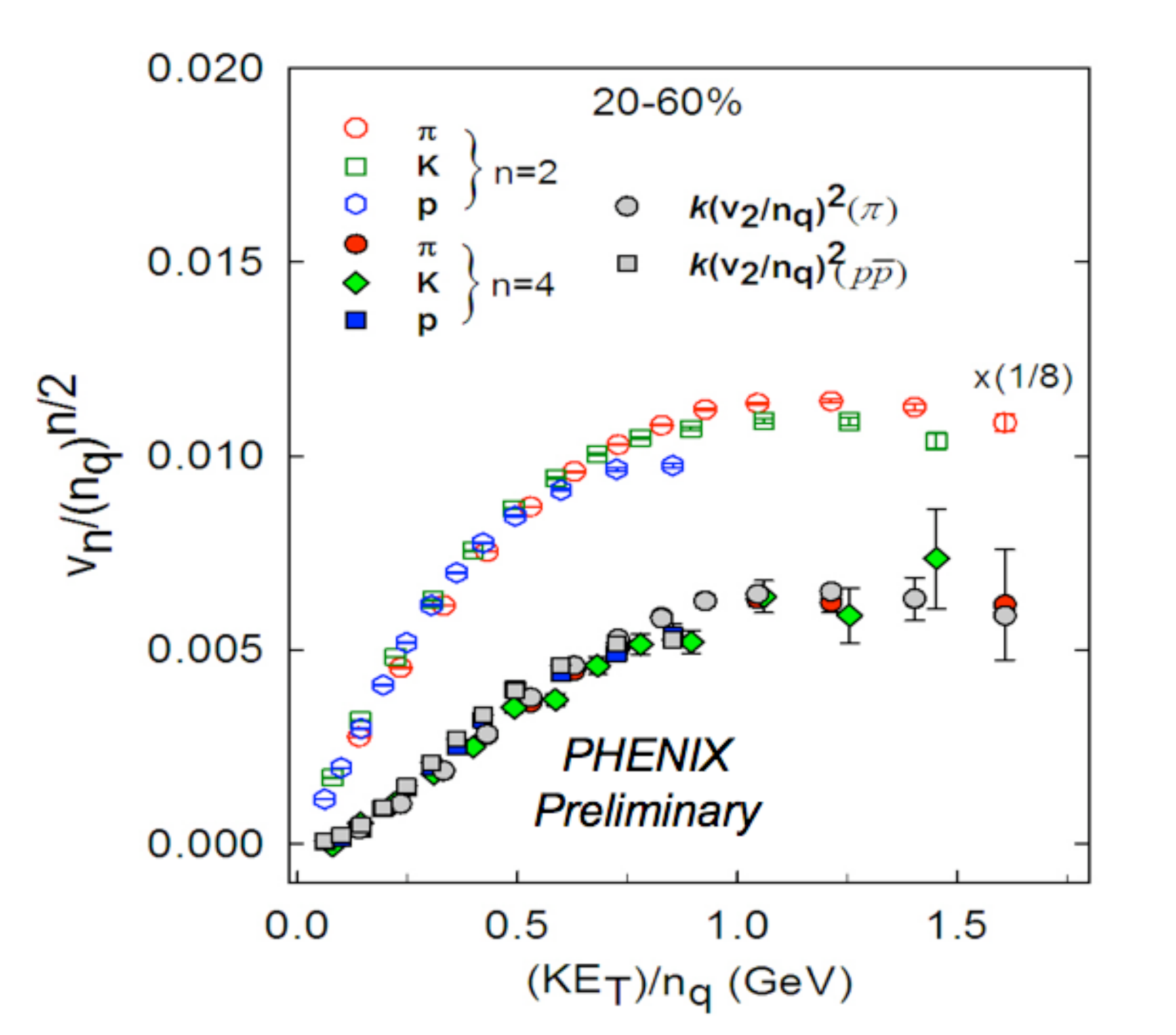}
\caption[]{(color online) Summary of scaling relations found for $v_2$ (open symbols) and $v_4$ (closed red, green and blue symbols) in Au+Au collisions for centralities $20-60\%$. The grey closed symbols show the scaling relation between $v_2$ and $v_4$, separately for pions and (anti-)protons.}
\label{v2v4scale}
\end{figure}

The observation of strong elliptic flow at RHIC was a crucial ingredient in establishing that the medium produced in heavy ion collisions at RHIC is a strongly coupled near-perfect liquid. The usefulness of azimuthal anisotropy studies is however far from exhausted, as the increased statistical reach of the data, combined with upgraded detectors, allow us to proceed with more differential studies of $v_2$ (the second Fourier component of the azimuthal distribution), and reach into the higher order flow coefficient $v_4$. Several scaling properties have been found for $v_2$\cite{PPG062}, namely that mesons and baryons scale separately as a function of the transverse kinetic energy $KE_T$, and that all hadrons collapse to a common curve when $v_2$ and $KE_T$ are both scaled by $n_q$, the number of constituent quarks, suggesting that the elliptic flow emerges early in the dynamical evolution of the system, while the medium is still described by partonic degrees of freedom. We have recently been able to verify that the same scaling applies also to $v_4$ (with $n_q^2$), and that $v_4/(v_2)^2$ is approximately constant, both of which are important steps in validating the hydrodynamical description of the QCD medium produced at RHIC. Figure \ref{v2v4scale} summarizes these results, and newer results not shown here \cite{AT} confirm that these scaling relations hold as a function of centrality even when studied in fine centrality bins. As has been the case with other probes, the added precision will allow us to develop analyses that are able to constrain model parameters and extract quantitative results on the transport coefficients that characterize the evolution of the system. 

Even though it still quantifies an azimuthal asymmetry, at higher $p_T$ $v_2$ can no longer be simply understood as originating from the geometry of the medium and its hydrodynamical evolution, as the high momentum partons that fragment into the particles populating this region of the spectrum are not expected to be a part of the expanding medium, even though they are affected by it through jet quenching. It is precisely the jet quenching mechanisms that are responsible for the non-zero $v_2$ observed at high $p_T$, since the different possible orientations of a jet with respect to the reaction plane result in different path lengths in the medium that the jet will have to travel through. 

\begin{figure}[t]
\centering
\includegraphics[width=0.98\textwidth]{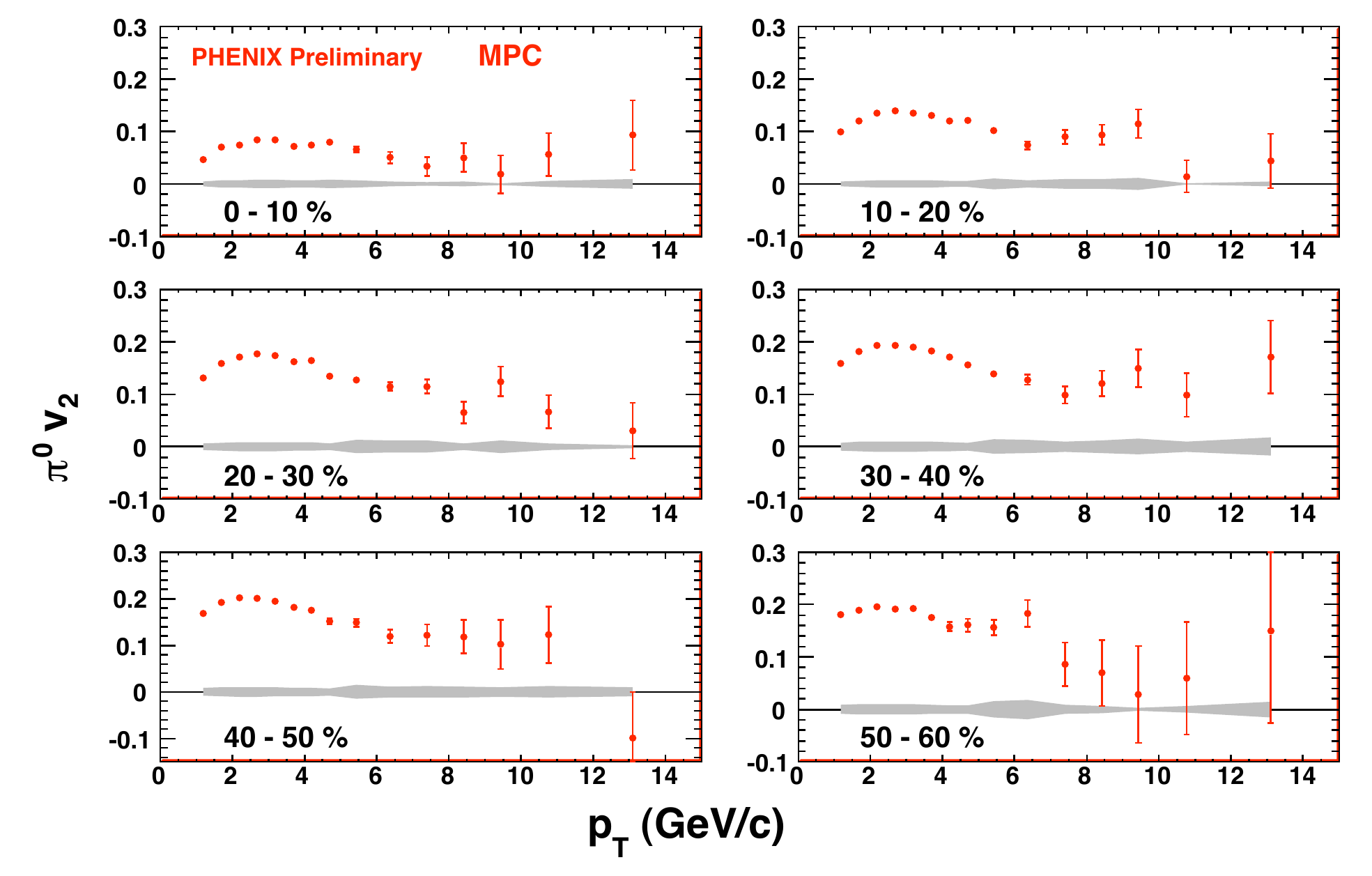}
\caption[]{Neutral pion $v_2$ in Au+Au collisions as a function of $p_T$ for different centrality bins, from most central (top left) to most peripheral (bottom right). The reaction plane used was estimated from the PHENIX Muon Piston Calorimeter (MPC), at $3 < |\eta| < 4$.}
\label{v2HighpT}
\end{figure}

Since Run-7 PHENIX has several new detectors available with full azimuthal coverage at different rapidity intervals that can be used to determine the reaction plane. These can be used to evaluate the effect of non-flow on the measured $v_2$, in particular at high $p_T$, where most jet-induced non-flow effects are expected to be significant, in particular near mid-rapidity. Indeed it was verified \cite{Rui} that determining the reaction plane further away from mid-rapidity ($3 < |\eta|<4$) reduces the non-flow effects the most. Figure \ref{v2HighpT} shows the $v_2$ as a function of $p_T$ for $\pi^0$s, in centrality bins ranging from the $10\%$ most central to $50-60\%$ central. For all the centrality bins, at high $p_T$ the $v_2$ values are low but non-zero, and show little $p_T$ dependence.

The main observable used in the study of jet energy loss is the nuclear modification factor, $R_{AA}$, the ratio of the A+A single particle yield scaled by the average number of binary collisions in a given centrality bin, to the yield in p+p collisions. Despite the availability of results for the $\pi^0$ $R_{AA}$ centrality dependence all the way to transverse momenta of nearly 20~GeV/$c$ \cite{PPG080}, we are not yet able to distinguish between the various energy loss models that describe the data well, even though there are significant differences in the approaches and assumptions among the models. In order to be able to discriminate between the models an additional handle on the path length of the medium can be obtained by studying $R_{AA}$ as a function of the angle to the reaction plane, in addition to $p_T$ and centrality.

\begin{figure}[t]
\centering
\includegraphics[width=0.95\textwidth]{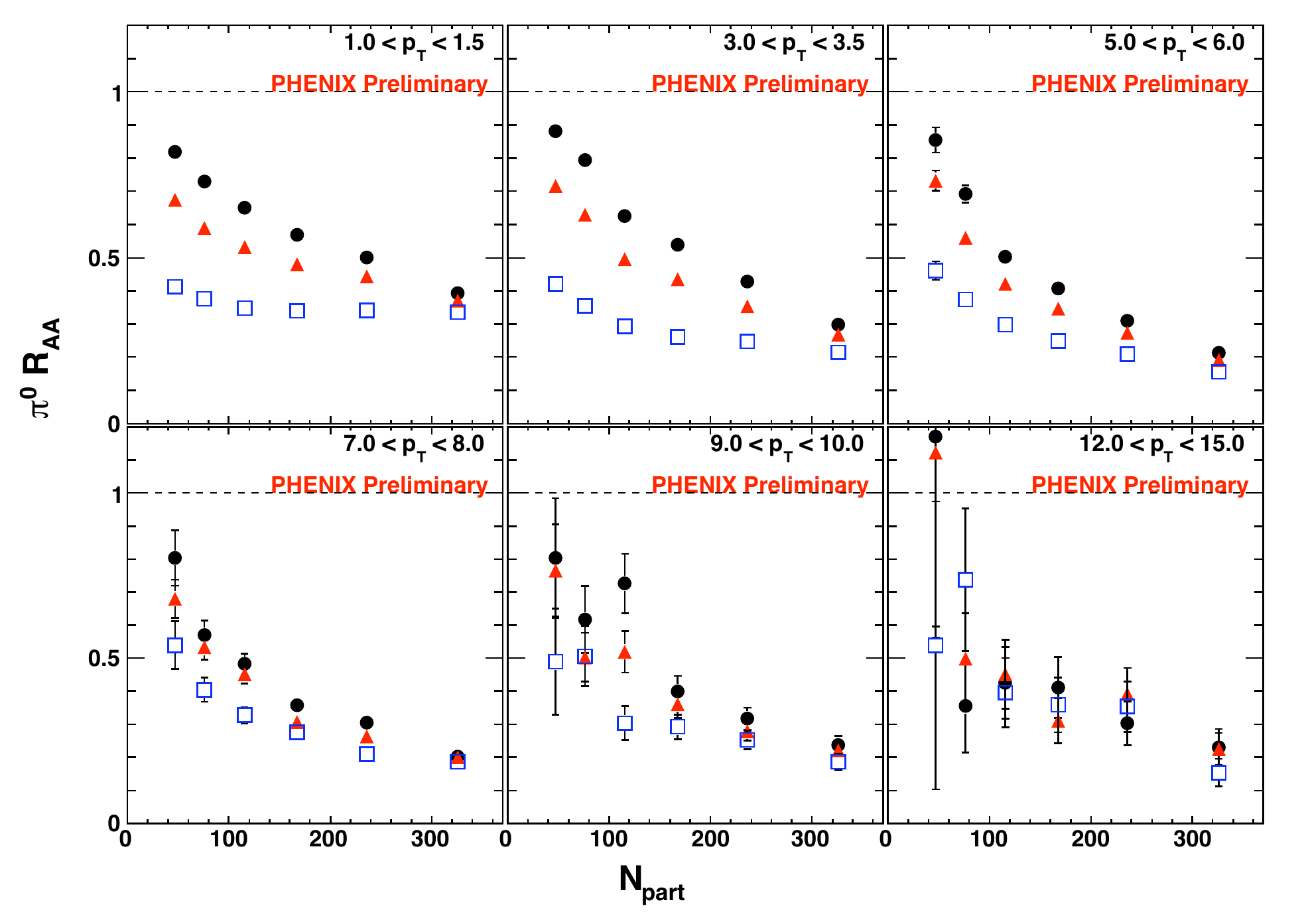}
\caption[]{(color online) $\pi^0$ $R_{AA}$ as a function of $N_{part}$ for several bins of $p_T$, as identified in each panel. The black circles correspond to $0^\circ < \Delta\phi <  15^\circ$ (in-plane), the red triangles to $30^\circ < \Delta \phi < 45^\circ$ (intermediate) and the open blue squares to $75^\circ < \Delta \phi < 90^\circ$ (out-of-plane).}
\label{inpoutp}
\end{figure}

Figure \ref{inpoutp} shows $R_{AA}$ as a function of the number of collision participants, $N_{part}$, for a sampling of bins of $p_T$ and three orientations with respect to the reaction plane: in-plane, out-of-plane, and intermediate. The main trends that can be seen were already identified in \cite{PPG092}, but the current preliminary result expands the reach from 10 to 15~GeV/$c$ in $p_T$. These trends are: at lower $p_T$, the out-of-plane $R_{AA}$ is nearly independent of centrality, for $N_{part}>100$; the out-of-plane $R_{AA}$ on the other hand shows a strong centrality dependence, dropping by about a factor of 2 from peripheral to central collisions; and for the highest $p_T$ the three curves seem to be approaching convergence. Related results on $R_{AA}(p_T)$ \cite{Rui} for in- and out-of-plane angles can be compared to existing energy loss models \cite{Bass}, which seem to have some difficulty in matching these data, despite their success in describing the angle-integrated $R_{AA}(p_T)$.

\section{Two-Particle Correlations}

The typical image conveyed by two-particle (most commonly, two-hadron) azimuthal correlations is that of two peaks centered at azimuthal angle differences $\Delta \phi$ of approximately zero and $\pi$, usually referred to as the near and away-side peaks, respectively. In p+p collisions this is understood as resulting from a high momentum interaction that produces two back-to-back jets, but in Au+Au collisions the scenario is complicated by the presence of the medium that both jets have to traverse as they evolve and hadronize, and this complication is what makes two-particle collisions an excellent tool to study the mechanisms of jet energy loss and the response of the medium to the energy that is left behind by the quenched jets. In addition to the original observation of the disappearance of the away-side jet, other features observed in di-hadron correlations are under active study and discussion, such as the ``ridge", an enhancement of the near-side peak along the $\Delta \eta$ direction, the ``shoulder" region where a double peak structure is seen, around $\Delta \phi \sim \pi \pm1$, and the ``head" region around $\Delta \phi \sim \pi$ where either the away-side peak disappears or a punch-through peak emerges, depending on the associated particle momenta. 

As discussed previously for single particle measurements, studying the geometric dependence of energy loss can be achieved by performing measurements that are oriented at different angles relative to the reaction plane. That way we can control both the length of the medium that is seen by the jet and also how (or whether) the medium response can be disentangled from the hydrodynamical flow effects. By using different $p_T$ ranges for the trigger and associated hadrons these two different questions can be explored by the same type of analysis \cite{Wolf}.

\begin{figure}[t]
\centering
\includegraphics[width=0.99\textwidth]{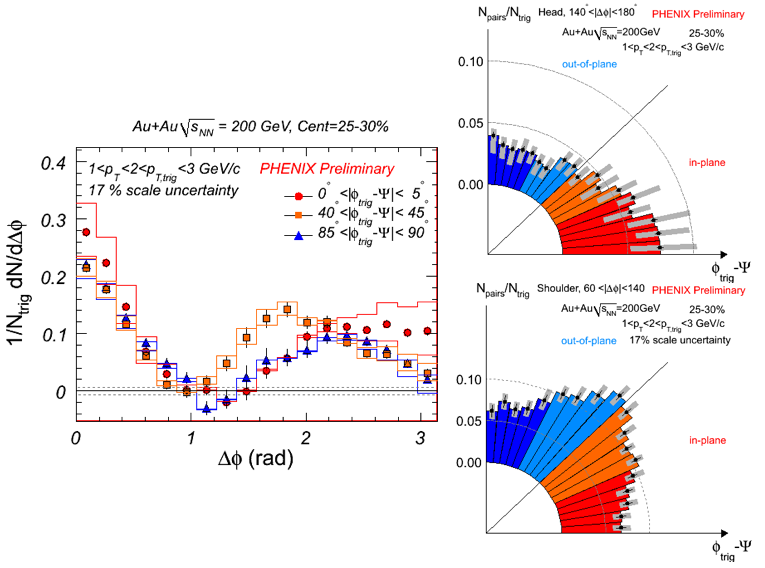}
\caption[]{(color online) Left: Per trigger jet functions for mid-central Au+Au collisions, with $1< p_{T,assoc} < 2 < p_{T,trig} <3$~GeV/$c$. Right: Per trigger yields for the head: $140^\circ<|\Delta \phi|<180^\circ$ (top), and shoulder: $60^\circ<|\Delta \phi|<140^\circ$(bottom) regions, going from in-plane (red) to out-of-plane (dark blue).}
\label{RPcorrLowpT}
\end{figure}

PHENIX has results on correlations of two charged hadrons relative to the reaction plane, at intermediate $p_T$, in the ranges of $2 < p_T < 3$~GeV/$c$ for the trigger hadron, and $1 < p_T < 2$~GeV/$c$ for the associate hadron, aiming to systematically study the geometry dependence of the medium response and constrain models that describe possible medium modification mechanisms. Figure~\ref{RPcorrLowpT} shows the results for an intermediate centrality bin, and three angular orientations of the trigger particle with respect to the reaction plane. In the head region of the away-side, the in-plane (red) and out-of-plane (blue) yields show the most pronounced difference, with a peak only present for the in-plane angle. Despite this yield difference near $\Delta \phi$ of $\pi$, the in-plane and out-of-plane jet functions show very similar widths in the away-side. The intermediate angle (yellow, near $45^\circ$) exhibits yet another trend, with the yield in the head region being almost as suppressed as for the out-of-plane angle, but with a peak emerging near $\Delta \phi \sim \pi - 1.3$. The right panel of the figure, showing the integrated yields in the head and shoulder regions as a function of the angle, very clearly showcases all of the features of the jet functions. 

By studying the reaction-plane angle dependence of the away-side yields of two-particle pairs where both members have relatively high transverse momenta, we can distinguish between two general models of their production. In the first scenario, the away-side is composed of punch-through jets, that lost part of their energy in the medium but still made it out with a significant portion of their initial momentum. The angle between the trigger particle and the reaction plane controls the path length of the medium that the punch-through jet sees, and therefore we expect lower $\phi$ (shorter path length) to correspond to higher away-side yields, and higher angles (longer path length) to lower yields. In the other production picture, most of the medium is a central opaque core that fully suppresses jets, and therefore all the observed pairs must be produced near the surface, and tangentially to the medium. This type of tangential production will result in the opposite dependence of the away-side yields with the reaction plane angle. These two simple opposing scenarios may however be complementary, if for instance the opaque core is only present near the center of the medium, or not completely opaque. 

\begin{figure}[t]
\centering
\includegraphics[width=0.48\textwidth]{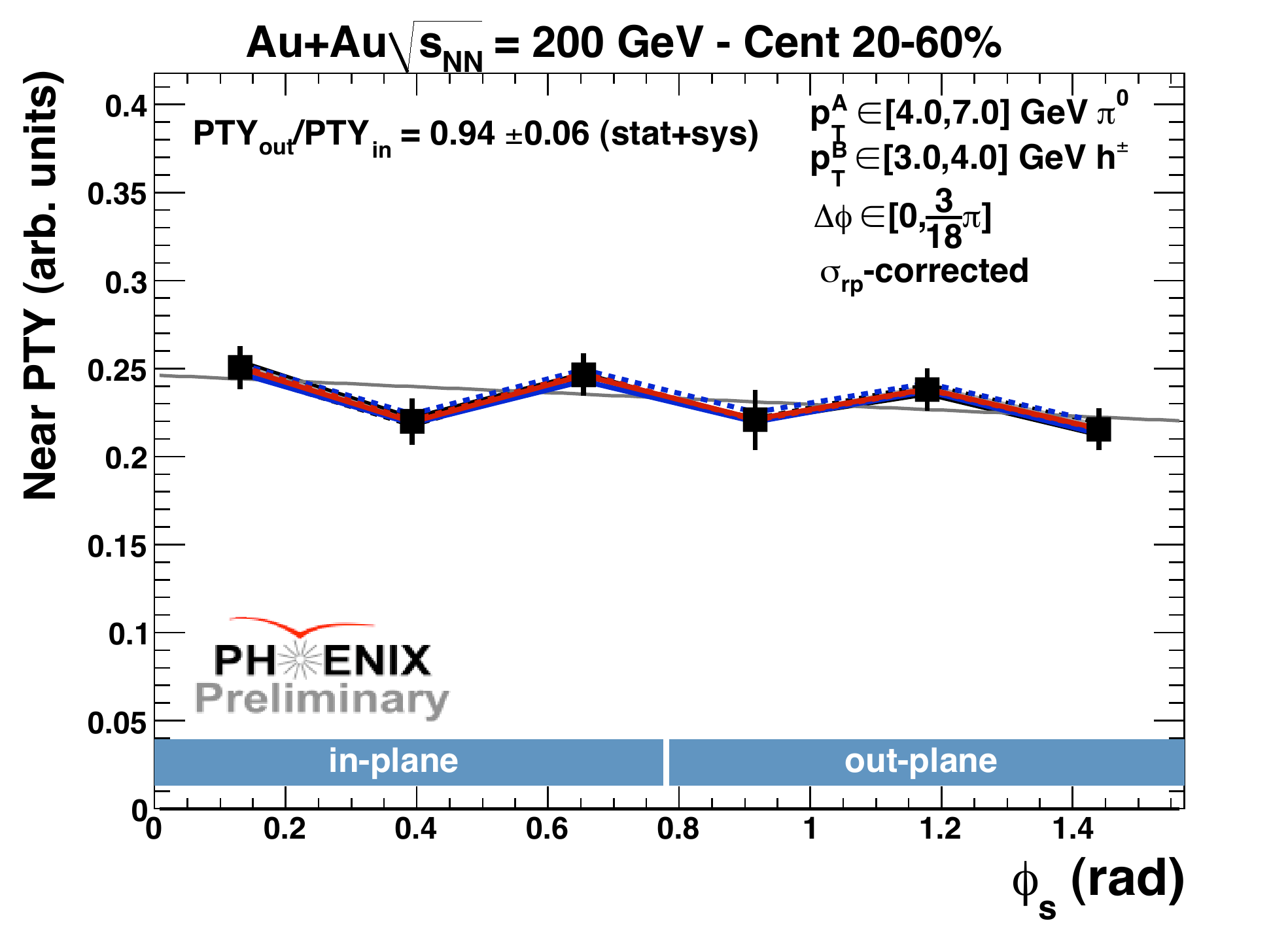}
\includegraphics[width=0.48\textwidth]{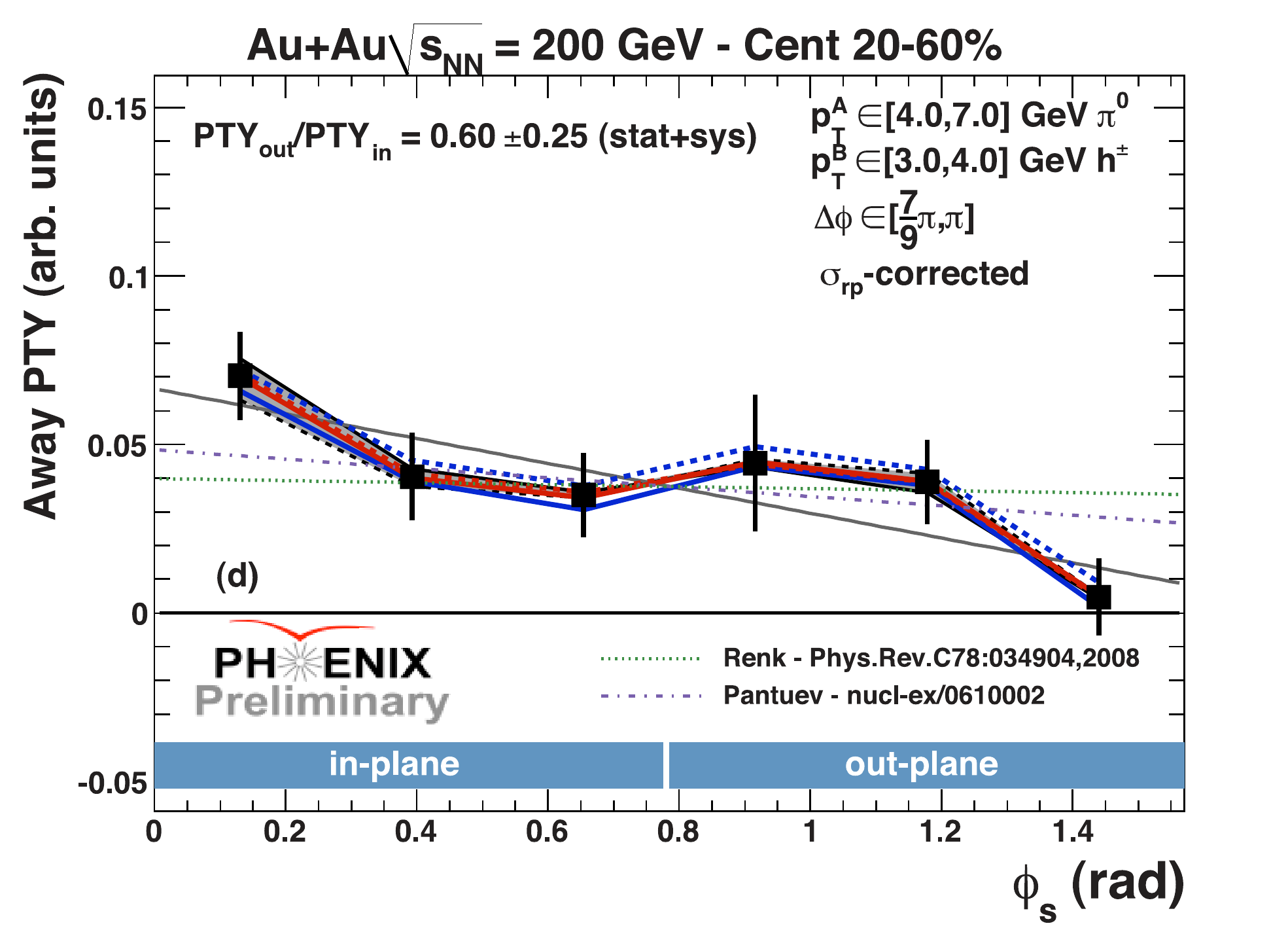}
\caption[]{Near (left) and away-side (right) integrated per trigger yields as a function of angle $\phi_s$ between the trigger and the reaction plane, for Au+Au collisions in the centrality range $20-60\%$. The solid line is a simple linear fit, dashed lines are from references indicated in the figure.}
\label{RPcorrHighpT}
\end{figure}

Figure~\ref{RPcorrHighpT} shows the near and away-side per trigger yields as a function of $\phi_s = \phi_{trig} - \Psi_{RP}$, for the centrality range $20-60\%$. The near-side yields are essentially flat when going from in-plane to out-of-plane, but in the away-side yields a steep dependence is observed, with yields being highest in-plane, as expected from the penetrating production picture. For more central collisions, the error bars do not allow a clear conclusion as to whether the same trend is observed \cite{McCumber}.

\section{$\gamma-$Jet and Full Jet Reconstruction}

In the search for quantitative handles on energy loss, correlation measurements between direct photons and jets are seen as the ideal tomographic probe, as the jet's initial momentum is approximately balanced by the momentum of the photon, which is expected to emerge from the medium unaffected. PHENIX has indeed verified that the nuclear modification factor for direct photons is very close to unity up to transverse momenta of at least 14~GeV/$c$. The $\gamma$-jet measurements are done by correlating inclusive photons with high-momentum hadrons, and statistically subtracting the decay photon component.  

The first PHENIX results on direct $\gamma$-jet measurements have recently been submitted for publication \cite{PPG090}, and show that the away side yield is suppressed relative to that observed in p+p collisions, to a level that is consistent with that seen for single particles. The jet fragmentation function can be expressed as: 
\begin{equation}
D_q (z_T) = \frac{1}{N_{evt}}\frac{dN(z_T)}{dz_T}
\end{equation}
where $z_T = p_T^{h}/p_T^{jet}$, which can be approximated by $p_T^{h}/p_T^{\gamma}$. Its modification can be observed through deviations from the exponential $z_T$ scaling observed in p+p collisions. 

\begin{figure}[t]
\centering
\includegraphics[width=0.98\textwidth]{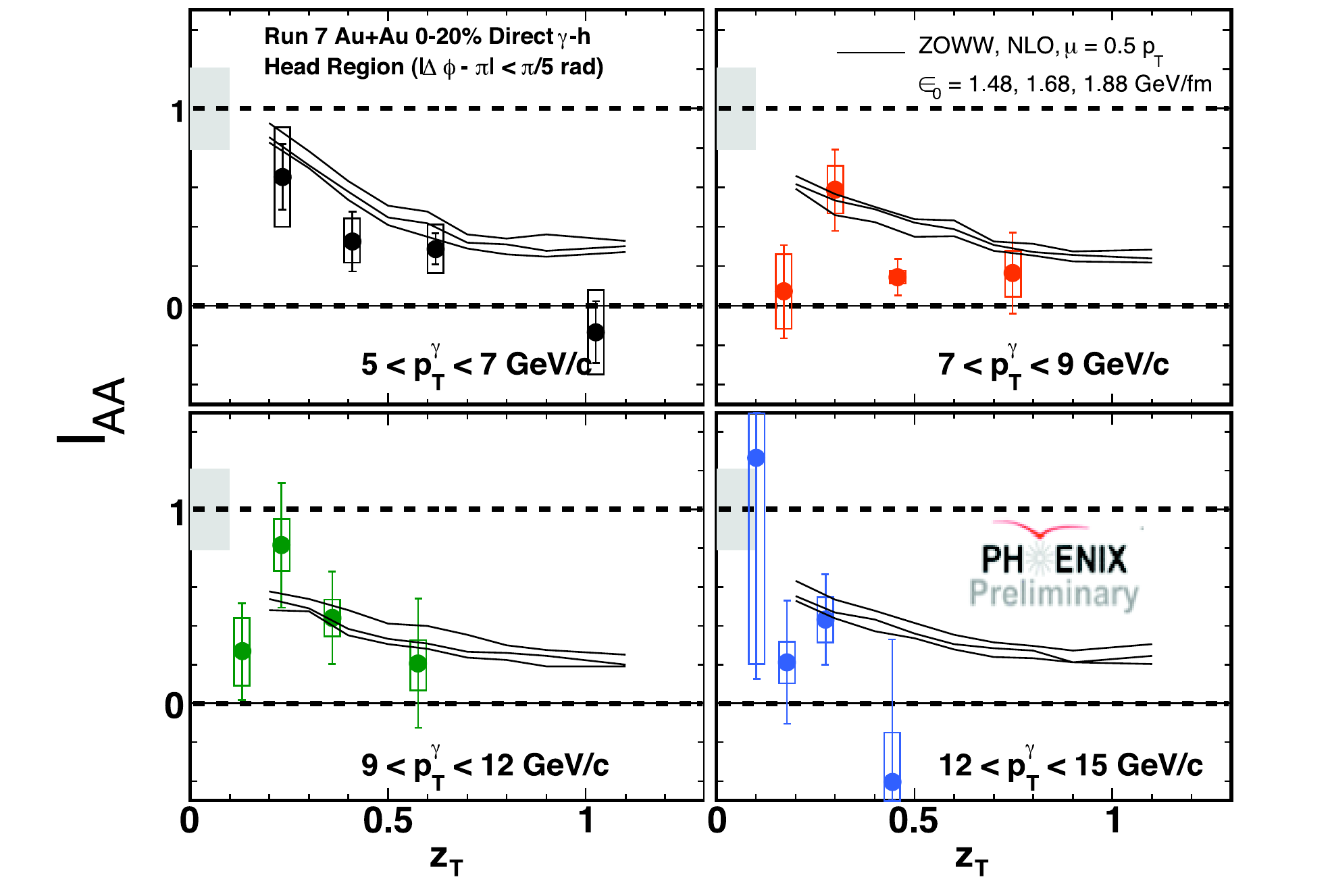}
\caption[]{Nuclear modification of the away-side two-particle yield, $I_{AA}$, in the head region, as a function of $z_T$ for several bins of the photon's transverse momentum. Also plotted are theory curves from \cite{zoww}.}
\label{gammajetIAA}
\end{figure}

Preliminary higher statistics results from Run-7 are now available, expanding the measurement to higher $p_T^{hadron}$ and providing a clearer picture of the modification of fragmentation function. The result for the nuclear modification of the away-side yield, $I_{AA}$ (ratio of Au+Au to p+p per trigger yields), for the 20\% most central Au+Au collisions, is shown in figure \ref{gammajetIAA}, as a function of $z_T$. The observed suppression appears to increase with $z_T$, a trend which is also seen in the theory curves shown\cite{zoww}. This is understood by considering the geometry involved: for low $z_T$, the corresponding hadrons originate from a wider range of depths in the medium, whereas at higher $z_T$ they suffer from the surface bias inherent in the requirement of observing a high-momentum hadron. Other model comparisons are presented and discussed in \cite{Megan}.

A comparison of the slopes of the away side yield as a function of $z_T$ in central Au+Au and p+p collisions \cite{Megan} shows the slope for Au+Au ($9.49\pm 1.37$) is 1.3$\sigma$ higher than the one measured for p+p ($6.89 \pm 0.64$), suggesting that in central Au+Au collisions the quark fragmentation function is modified, in addition to being suppressed. Both sets of points are well fitted by the exponential functions used.

A new approach in the study of jets and energy loss is emerging at RHIC, with the application of algorithms for full jet reconstruction in heavy ion collisions. PHENIX presented preliminary results on jet reconstruction in p+p and Cu+Cu using a Gaussian filter algorithm \cite{Lai}, including reconstructed jet spectra and di-jet angular correlations \cite{YueShi}. Further results using this method are expected soon, and will be an essential element in our search for a complete picture of jet physics in heavy ion collisions. 

\section{Summary}

The PHENIX collaboration has performed extensive measurements in A+A collisions, continuing to widen our knowledge and understanding of the hot and dense medium created at RHIC. Detailed studies of anisotropic flow validate the hydrodynamical description of the initial evolution of the bulk medium, and show that the flow is established at the partonic level. Azimuthal dependences in various measurements at high-$p_T$, both from single particles and correlations, are uncovering the details of jet quenching mechanisms in the medium. Finally, $\gamma-$jet and full jet reconstruction studies bring us closer to true tomography at RHIC. It is worthwhile noting that while several of the observables discussed here have already reached a level of precision and maturity that allows them to be used to constrain theoretical models and work towards a quantitative description of the medium we are exploring, others are just starting to develop, and will require the upcoming RHIC luminosity and experimental upgrades to fully accomplish their goals.

\end{document}